\documentclass[prx,twocolumn]{revtex4}  

\usepackage{amssymb}
\usepackage{amsmath}
\usepackage{graphicx}
\usepackage{bbm}
\usepackage{color}
\usepackage{comment}
\usepackage{gensymb}
\usepackage{gensymb}
\definecolor{blue}{rgb}{0,0,1}
\definecolor{grey}{rgb}{0.6,0.6,0.6}

\newcommand{\bra}[1]{\langle #1 |}
\newcommand{\ket}[1]{| #1 \rangle}

\begin{document}

\title{Quantum state transfer via acoustic edge states in a 2D optomechanical array}
\author{Marc-Antoine Lemonde$^1$, Vittorio Peano$^2$, Peter Rabl$^3$, Dimitris G. Angelakis$^{1,4}$}
\affiliation{$^1$ Centre for Quantum Technologies, National University of Singapore, 3 Science Drive 2, Singapore 117543, Singapore}
\affiliation{$^2$ Max Planck Institute for the Science of Light, Staudtstrasse 2, 91058 Erlangen, Germany}
\affiliation{$^3$ Vienna Center for Quantum Science and Technology, Atominstitut, TU Wien, 1040 Vienna, Austria}
\affiliation{$^4$ School of Electrical and Computer Engineering, Technical University of Crete, Chania, Crete, 73100, Greece}

\date{\today}

\begin{abstract}
We propose a novel hybrid platform where solid-state spin qubits are coupled to the acoustic modes of a two-dimensional array of optomechanical nano cavities. 
Previous studies of coupled optomechanical cavities have shown that in the presence of strong optical driving fields, the interplay between the photon-phonon interaction and their respective inter-cavity hopping allows the generation of topological phases of sound and light. 
In particular, the mechanical modes can enter a Chern insulator phase where the time-reversal symmetry is broken.
In this context, we exploit the robust acoustic edge states as a chiral phononic waveguide and describe a state transfer protocol between spin qubits located in distant cavities.
We analyze the performance of this protocol as a function of the relevant system parameters and show that a high-fidelity and purely unidirectional quantum state transfer can be implemented under experimentally realistic conditions. 
As a specific example, we discuss the implementation of such topological quantum networks in diamond based optomechanical crystals where point defects such as silicon-vacancy centers couple to the chiral acoustic channel via strain.
\end{abstract}

\maketitle


\section{Introduction}

In recent years the efforts towards building scalable quantum information processing devices have reached unprecedented intensities. For this purpose, a number of physical platforms, such as superconducting circuits~\cite{Wendin2017}, cold atoms in optical lattices~\cite{Bloch2008, Gross2017}, trapped ions~\cite{Bruzewicz2019}, Rydberg atoms~\cite{Saffman2010} and defect centers in solids~\cite{Neumann2010, Weber2010, Yao2012, Cai2013, Wang2015}, are actively investigated.
In parallel, various strategies for implementing hybrid quantum systems are currently explored~\cite{Wallquist2009, Xiang2013, Kurizki2015}, 
with the long-term goal to combine the strengths of the different architectures and to mitigate system-specific weaknesses.
In this context, high-Q mechanical elements play a particularly important role for realizing coherent quantum interfaces~\cite{Rabl2010, Stannigel2010, Schuetz2015, Bochmann2013, Bagci2014, Andrews2014,Rueda2016,Tian2015} as they can be coupled efficiently to a large variety of other quantum systems~\cite{Treutlein2014} while being themselves only weakly affected by decoherence. Similar to optical fields, acoustic waves can be guided along coupled resonator arrays or continuous phononic waveguides~\cite{SafaviNaeini2011, Khanaliloo2015, Patel2017}, which can be used to implement chip-scale quantum networks where quantum information is distributed via individual propagating phonons~\cite{SafaviNaeini2011, Habraken2012, Gustafsson2014, Lemonde2018, Bienfait2019}.
In particular, such phononic quantum channels have been proposed to overcome the problem of coherently integrating a large number of electronic spin qubits associated with defect centers in diamond~\cite{Rabl2010, Bennett2013, Albrecht2013, Lee2017, Lemonde2018, Kuzyk2018, Li2019}.
However, being in its infancy, the control of acoustic waves on the quantum level still faces many challenges, which must be met both on an experimental and on a conceptual level. This includes, for example, the scattering of phonons along the channel by fabrication imperfections, but even more fundamentally, the ability to emit phonon wavepackets with a specified shape and direction, as a prerequisite for many quantum state transfer protocols~\cite{Cirac1997}.

In this work we propose and analyze an hybrid phononic quantum network, where spin qubits or other two-level systems (TLS) are coherently coupled to the chiral acoustic edge channels of a two dimensional (2D) optomechanical (OM) array. This architecture is motivated by the progress in engineering spin-phonon interactions in solid-state systems~\cite{MacQuarrie2013, Ovartchaiyapong2014, Barfuss2015, Golter2016, Meesala2016, Jahnke2015, Sohn2017, Meesala2018}, as well as in fabricating 2D OM crystals~\cite{Safavi2010, Gavartin2011, Safavi2014, Kalaee2016} with different geometries and band structures. 
In a previous work~\cite{Peano2015} it has been shown that  2D OM arrays can exhibit a rich set of topological phases of sound and light that can be fully explored by tuning in situ the optical driving of the cavities.
In particular, for weak OM interactions, the acoustic excitations are expected to enter a Chern insulator phase where chiral edge states propagate along the array boundaries. Thus this hybrid quantum system offers a platform to study rich physics emerging from the interplay between spins, mechanical and optical degrees of freedom in phases where time-reversal symmetry is broken.

As a first application for this setup we focus on the quantum-state transfer between TLS located in distant cavities via chiral acoustic edge channels. Compared to state transfer protocols in regular 1D phononic waveguides~\cite{SafaviNaeini2011, Habraken2012, Lemonde2018,Bienfait2019}, this platform offers the advantages of a unidirectional propagation~\cite{Cirac1997, Yao2013, Barik2018, Ozawa2019}, which is robust against local perturbations and where the direction can be controlled by external optical driving fields. While the basic protocol discussed in this work is very general, a naturally-suited system where these ideas can be implemented is an array of separated Silicon vacancy (SiV) centers in a diamond OM crystal. In this case, quantum information can be stored in the long-lived spin degrees of freedom of the SiV ground state~\cite{Goss1996, Hepp2014, HeppThesis, Sukachev2017, Becker2017}, where at low temperatures of $T\lesssim1$ K coherence times exceeding $T_2\sim 10$ ms have been demonstrated~\cite{Becker2017, Sukachev2017}. At the same time the orbital degrees of freedom of the defect allow strong and tunable coupling to vibrational modes, as recently discussed in Ref.~\cite{Lemonde2018}. Combined with the ability to design chiral acoustic channels via OM interactions this coherent spin-phonon interface offers many new tools to overcome fundamental challenges in phononic quantum network applications.


\begin{figure}[t]
	\begin{center}
	\includegraphics[width= 0.95\columnwidth]{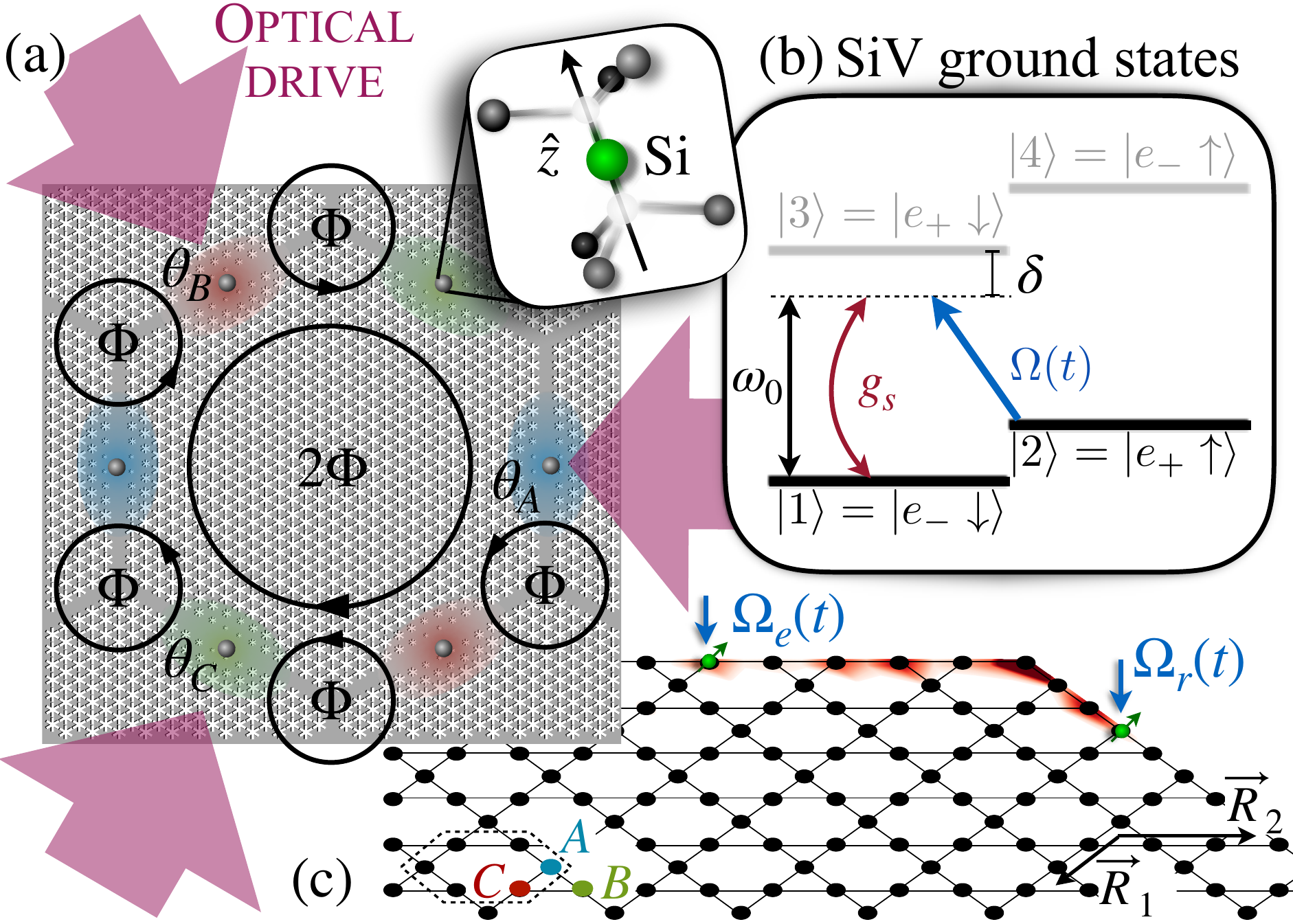}
	\end{center}
	\vspace{-0.5cm} 
	\caption {
Schematics of the 2D hybrid system and the state transfer protocol. (a) OM cavities engineered from smooth alterations of a snowflake-hole patterning in diamond. The cavities are arranged in a Kagome structure where the unit cells include a basis of three sites $s=\{A,B,C\}$. A TLS is embedded in every site. The phase pattern $\{\theta_s\}$ induced by the optical driving generates a flux $\Phi$ upon hopping anti-clockwise through a unit cell. (b) Energy structure of the SiV ground-state subspace where the two lowest-energy states $\ket 1$ and $\ket2$ form a long-lived spin qubit. A microwave drive $\Omega(t)$ couples opposite spin states while the strain associated with a single phonon couples  the orbital degrees of freedom $\ket{e_\pm}$ with strength $g_s$. The combination of the two processes leads to a tunable interaction between the spin states and phonons  of frequency $\sim \omega_0$. (c) State transfer between distant TLS via topologically protected chiral acoustic waves propagating along the boundaries of the structure. 
}
\label{Fig:Schema}
\end{figure}


\section{Model}

We consider a 2D array of OM cavities as depicted in Fig.~\ref{Fig:Schema}, where each lattice site contains a single TLS. The OM array can be realized, for example, in so-called ``snowflake'' structures~\cite{Safavi2010T}, where high-Q vibrational and optical modes are co-localized in regions of engineered defects created by smoothly varying the size of the patterned periodic snowflake holes [cf.~Fig.~\ref{Fig:Schema} (a)]. 

At each lattice site $j$, the variation in the index of refraction due to mechanical vibrations leads to a strong OM coupling that can be described by the standard OM Hamiltonian 
$\hat H_{\rm OM}/\hbar = \omega_M \hat b_j^\dag \hat b_j + \omega_c \hat a_j^\dag \hat a_j + g_0 \hat a_j^\dag \hat a_j (\hat b_j + \hat b_j^\dag)$~\cite{Aspelmeyer2014}.
Here $\hat a_j$ ($\hat b_j$) represents the annihilation operator of the photonic (phononic) mode of frequency $\omega_c$ ($\omega_M$) and $g_0$ is the OM coupling per photon. Due to the strong localization of both photons and phonons, this coupling can reach values of about $g_0\sim \, 250$ kHz~\cite{Safavi2014}, which we will assume for all the following estimates. 
The optical cavities are driven by a strong external laser field of frequency $\omega_L$, which drives each optical mode into a coherent state with amplitude $\alpha_j(t)=\sqrt{n_c}e^{i\theta_j}e^{-i\omega_L t}$, where $n_c\gg 1$ is the mean intracavity photon number. By redefining $a_j \rightarrow a_j+\alpha_j(t)$, the OM interactions can be linearized and in a frame rotating with $\omega_L$, the resulting  Hamiltonian for the whole 2D OM array is given by $(\hbar=1)$
\begin{align}
	\hat H_{\rm OMC} & = \sum_j ( \omega_M \hat b_j^\dag \hat b_j - \Delta \hat a_j^\dag \hat a_j + G e^{i\theta_j} \hat a_j^\dag \hat b_j + Ge^{-i\theta_j} \hat b_j^\dag \hat a_j ) \nonumber \\
	& \qquad + \sum_{<i,j>} (K \hat b_i^\dag \hat b_j + J \hat a_i^\dag \hat a_j + {\rm H.c.} ). \label{Eq:HOMC}
\end{align}
Here  $J>0$ ($K>0$) denotes the nearest-neighbor photon (phonon) hopping rate and $\Delta = \omega_L - \omega_C < 0$ is the detuning between the cavities and the drive. In Eq.~\eqref{Eq:HOMC}, $G =  g_0 \sqrt{n_c}$ is the linear OM coupling, which is enhanced by the number of photons in a cavity. 
The positive hopping rates considered here are in contrast to the model proposed in Ref.~\cite{Peano2015} and, as described below, lead to qualitatively different scenarios.
At this stage, we consider all parameters identical throughout the lattice except for the driving phases $\theta_j$. Note that in Eq.~\eqref{Eq:HOMC}, we made an additional rotating-wave approximation by neglecting processes that do not conserve the number of excitations ($\sim G e^{-i\theta_j} a_j b_j + {\rm H.c.}$). The validity of this approximation is discussed in Appendix~\ref{App:RWA}.


\begin{figure}
	\begin{center}
	\includegraphics[width= 0.7\columnwidth]{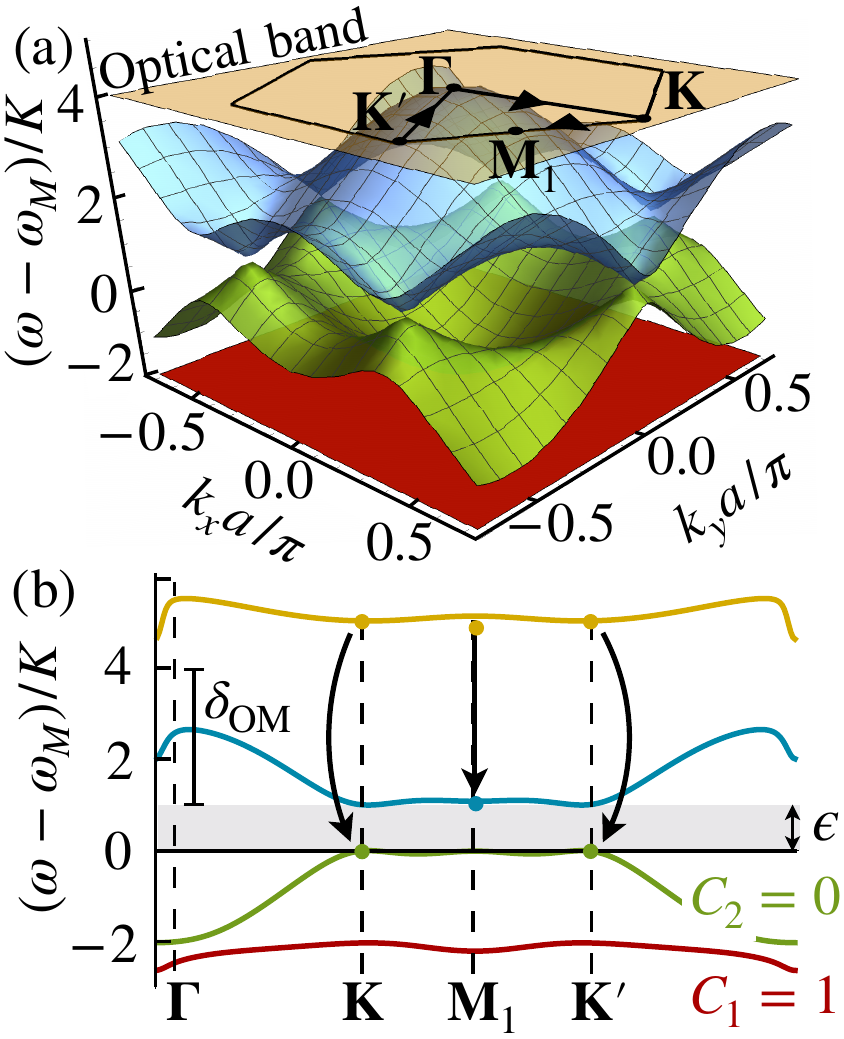}
	\end{center}
	\vspace{-0.5cm} 
	\caption {
OM band structures. (a) Dispersion relation of the non-interacting OM system ($G  = 0$), where the three lowest-energy bands represent mechanical modes while the  highest-energy band is optical in nature. The drive detuning here is $\Delta = -\omega_M - 2J - 4K$, which leads to $\delta_{\rm OM} = 3K$, and $J/K = 200$. 
(b) Dispersion relation along the path ${\bf \Gamma} \rightarrow {\bf K} \rightarrow {\bf M}_1 \rightarrow {\bf K'} \rightarrow {\bf \Gamma}$ in the Brillouin zone [depicted in (a)] for $G = 2K$, $\delta_{\rm OM} = 3K$ and $J/K = 200$. 
The black arrows indicate the dominant angular-momentum conserving OM interactions responsible for the gap.
The Chern numbers $C_{l}$ associated with the two lowest-energy mechanical bands are shown.}
\label{Fig:DispRel}
\end{figure}

In addition to the localized optical and mechanical modes, we consider a TLS embedded in each sites of the array, which is coupled to the acoustic vibrations via strain. We model the interaction by a Jaynes-Cumming coupling with time-dependent strength $g_{\rm sp}(t)$, such that the effective Hamiltonian describing the full hybrid 2D array reads
\begin{equation}
\hat H = \hat H_{\rm OMC} + \sum_j\frac{\omega_0}{2}\hat \sigma^{(j)}_z + \sum_j  \left[ g^{(j)}_{\rm sp}(t)\hat  \sigma_+^{(j)} \hat b_j   + {\rm H.c.} \right].
\label{Eq:Hnetwork}
\end{equation}
Here $\omega_0$ is the transition frequency of the TLS, $\hat \sigma_z$ is the usual Pauli-Z operator and $\hat \sigma^{(j)}_- = \hat \sigma_+^{(j)\dag}$ destroys a spin excitation in site $j$. 

While the spin-phonon coupling assumed in Hamiltonian \eqref{Eq:Hnetwork} is very generic and could be realized with various types of TLS~\cite{Habraken2012,Schuetz2015,Treutlein2014}, we explicitly  consider the example of SiV centers in diamond in our following analysis.  As depicted in Fig.~\ref{Fig:Schema}, the electronic  ground-state manifold of this center consists of two long-lived spin states denoted by $|1\rangle$ and $|2\rangle$, which can be coupled to a mechanical vibrational mode via a microwave assisted Raman process involving the excited state $|3\rangle$. 
More precisely, the strength of the time-dependent spin-phonon coupling $g_{\rm sp}(t)=\Omega(t) g_s/\delta$ and the qubit frequency $\omega_0$ can be externally tuned via the microwave drive amplitude $\Omega(t)$ and detuning $\delta$ compared to the state $\ket3$, respectively. Here $g_s$ is  the intrinsic strain coupling between the state $\ket1$ and $\ket3$.
Further details about SiV defects and their strain coupling are given in Appendix~\ref{App:SiV}.

\section{Acoustic edge channels} 

The main purpose of considering a 2D OM array instead of a simple 1D phononic waveguide is to use the OM interaction for engineering topologically protected acoustic edge channels, along which phonon propagation becomes unidirectional and immune against local disorder.  As first proposed in Ref.~\cite{Peano2015}, such a scenario can be achieved by imposing a non-trivial pattern of the driving phases $\theta_j$, which mimics the presence of a strong effective magnetic field.  Similar to electronic systems in real magnetic fields, the resulting bandstructure of the OM crystal may then exhibit topologically protected bands with a non-trivial Chern number, which for a finite system are associated with left- or right-propagating edge modes. 
In contrast to Ref.~\cite{Peano2015}, we here consider a different band structure which leads to much larger topological gaps in presence of weaker optical driving power.

\subsection{Topological phases of sound in an OM Kagome lattice}
While chiral acoustic edge channels can be implemented with various different OM lattice geometries, we here exclusively focus on the Kagome lattice for which topological phases of sound and light have already been described in Ref.~\cite{Peano2015}.  The Kagome crystal structure (see Fig.~\ref{Fig:Schema}) is defined by a triangular Bravais lattice spawned by the unit vectors $\{\vec R_1 = -(1,\sqrt3)a$, $\vec R_2 = (2,0)a\}$ and a three-cavity basis given by $\{\vec r_A = (0,\sqrt3/2)a$, $\vec r_B = -(1/2,\sqrt3/2)a$, $\vec r_C = (1/2,-\sqrt3/2)a\}$. Here $a$ is the distance between two adjacent cavities and $\{A,B,C\}$ refer to the different cavities within a unit cell. 
This structure possesses the full $\mathcal C_{6v}$ symmetry of the corresponding Bravais lattice.

In absence of the external driving fields, i.e.~$G=0$, the OM crystal system is time-reversal symmetric and  contains six energy bands. The three acoustic (optical) bands are centered around $\omega_M$ ($-\Delta$) and have a total width of  $6K$ (-$6J$). A zoom in of the non-interacting band structure in a spectral range that includes all acoustic bands but exclude far detuned optical modes is shown in Fig.~\ref{Fig:DispRel} (a).
We see that the ${\cal C}_6$ and time-reversal symmetries impose essential degeneracy at the high-symmetry points of the Brillouin zone, i.e.~${\bf K} = (2\pi/3a,0)$ and ${\bf K'} = (\pi/3a,\pi/\sqrt 3a)$, where Dirac cones are formed, and at ${\bf \Gamma} = (0,0)$, where a quadratic band-crossing point appears. Importantly, one of the optical (mechanical) bands is flat. This feature reflects the existence of localized normal modes describing a standing wave where the six cavities  along the edges of the same Wigner-Seitz cell are excited with equal amplitude but alternating  sign. More details about the diagonalization of the OM Hamiltonian in the quasi momentum space are given in Appendix~\ref{App:Hk}.

For finite driving of the OM cavities ($G\neq0$) the acoustic and photonic bands hybridize. Following Ref. \cite{Peano2015}, we choose the pattern of phases $\Delta\theta = \theta_B-\theta_A = \theta_C - \theta_B = \theta_A-\theta_C = \pm2\pi/3$ for every unit cells. In other words, we consider a driving of one of the optical modes at the ${\bf \Gamma}$ point. Such phase pattern can  be  generated by simply using three optical drives pointing at $120\degree$ angle from each others~\cite{Peano2015} [cf.~Fig.~\ref{Fig:Schema} (a)]. Most importantly,  it  breaks the time-reversal symmetry without breaking the spatial symmetries of the Kagome lattice and, thus, lifts the essential degeneracies giving rise to topological band gaps  \cite{Haldane1988}.


\begin{figure}
	\begin{center}
	\includegraphics[width= 0.7\columnwidth]{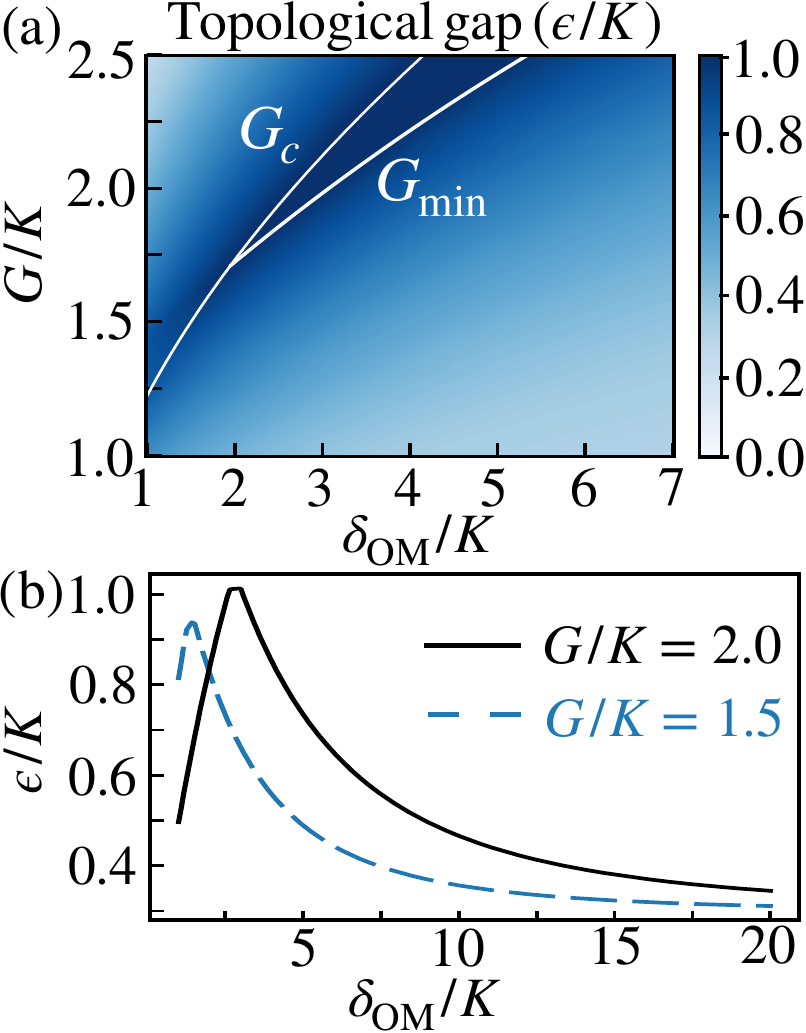}
	\end{center}
	\vspace{-0.5cm} 
	\caption {
Topological gap. (a) Topological gap size $\epsilon$ as a function of $G$ and the detuning $\delta_{\rm OM}$ for $J/K = 200$. The upper white line represents $G_c$ [cf.~Eq.~\eqref{Eq:Gc}] and the bottom one shows the minimal OM coupling $G_{\rm min}$ for which $\epsilon$ reaches $K$.
(b) Gap size as a function of $\delta_{\rm OM}$ for $G/K = 1.5$ and $G/K = 2.0$ with $J/K = 200$.}
\label{Fig:TopGap}
\end{figure}

	
\begin{figure*}[t]
	\begin{center}
	\includegraphics[width= 0.95\textwidth]{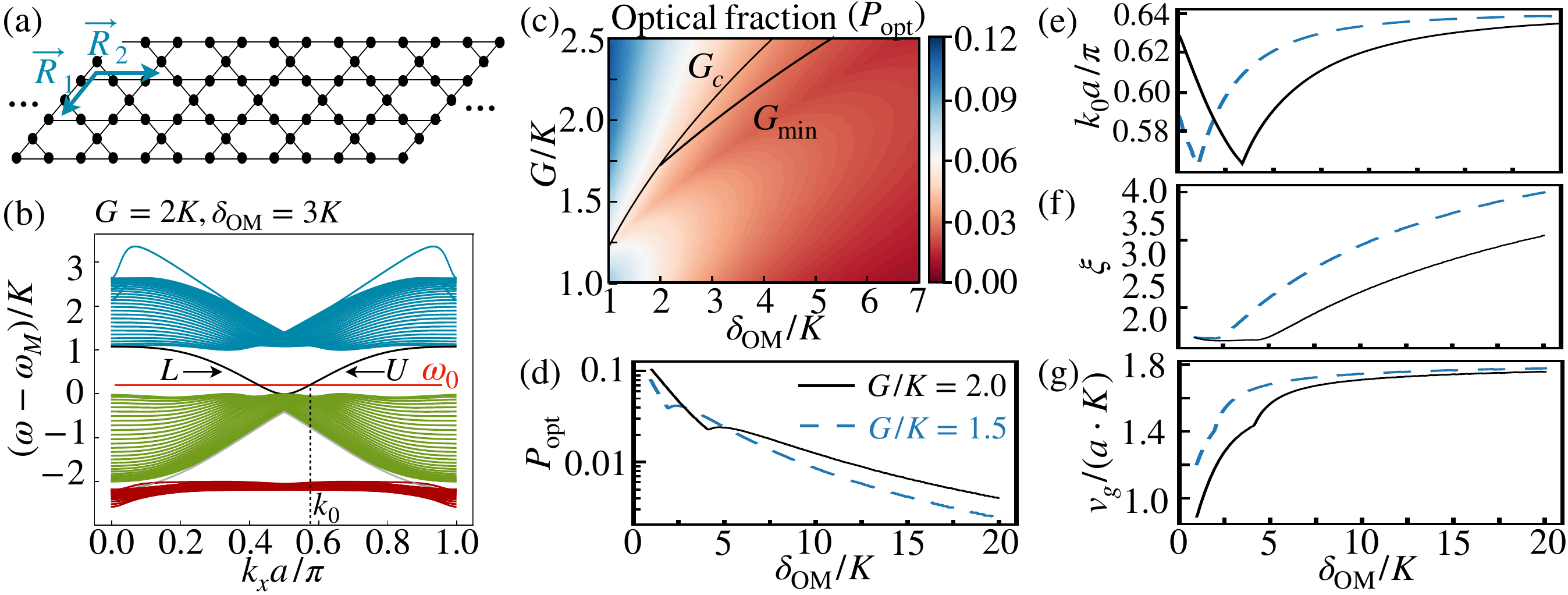}
	\end{center}
	\vspace{-0.5cm} 
	\caption {Topologically protected edge states. (a) Schematic of an infinite 1D stripe along $x$ with straight boundaries. (b) An example of the dispersion relation of the OM modes of the infinite stripe with $N_y = 21$ unit cells along $\vec R_1$, $G=2K$ and $\delta_{\rm OM} =3K$. The edge-state frequencies $\omega_E(k_x)$ are highlighted in black and the resonance frequency of the TLS, $\omega_0$, is shown in red. The edge mode propagating on the upper (lower) edge is labeled $U$ ($L$).
	(c) and (d) Optical fraction of the edge state evaluated at $k_0$ [shown in panel (e)] as a function of $G$ and $\delta_{\rm OM}$. The black lines in (c) indicate the value of $G_c$ and the value of $G_{\rm min}$.
	(f) Penetration depth $\xi(k_0)$ of the edge states as a function of $\delta_{\rm OM}$ for $G/K= 1.5$ and $G/K = 2.0$. 
	(g) Group velocity $v_g$ of the edge state propagating along the upper boundary ($\pi/2 < k_xa \leq \pi$) evaluated at $k_0$ as a function of $\delta_{\rm OM}$ for $G/K= 1.5$ and $G/K = 2.0$.
	In all figures, the photon hoping is $J/K = 200$.
}
\label{Fig:EdgeState}
\end{figure*}


Here, we focus on the weak OM coupling limit where the detuning of all optical modes is larger than the OM coupling, i.e.~$G\ll |\omega_M+\Delta+2J|$, such that all the excitations are still almost phononic or optical in nature. In this regime, the existence of topologically non-trivial phases for sound can be understood from the fact that phonons can also hop to neighboring lattice sites through virtual optical excitations. In the conceptually simplest setting where the optical bandwidth is small compared to the detuning, $J\ll |\omega_M+\Delta|$, this hopping is restricted to nearest neighbors and has an amplitude
\begin{equation}
K^{\rm opt}_{ij}\simeq \frac{G^2 J}{(\omega_M + \Delta)^2} e^{\pm i\Delta\theta_{ij}}.
\end{equation}
Since the resulting overall phonon hopping amplitude, $K_{ij}^{\rm eff} = K + K_{ij}^{\rm opt}$, then becomes a complex quantity, a phonon moving anti-clockwise around a crystal unit cell basis (i.e.~$A\rightarrow B\rightarrow C\rightarrow A$) acquires a phase $\Phi = \pm 3\arctan \frac{\sqrt3 JG^2}{JG^2 - 2K(\Delta + \omega_M)^2}$. This is reminiscent of a charged particle moving in a magnetic field where $\Phi$ represents the normalized magnetic flux encircled by the three cavities of the basis~\cite{Thouless1982}. 
Note that the total flux within a Bravais unit cell is zero as moving anti-clockwise over an hexagonal path leads to a phase shift of $-2\Phi$ [cf.~Fig.~\ref{Fig:Schema} (a)], simulating what is known as the anomalous quantum Hall effect~\cite{Haldane1988}.  In a more realistic situation, as considered in this work, the optical hopping rate is larger than the detuning, i.e.~$J\gtrsim |\omega_M+\Delta|$. In this case,  the same intuition holds but the optically-induced phonon hopping becomes longer-ranged and one must resort to a full numerical evaluation of the band structure, as exemplified in Fig.~\ref{Fig:DispRel} (b).
Note that in this same limit with $K\ll J$, the corresponding flux $\Phi_{\rm opt}$ experienced by the light field remains negligible, thus suppressing any non-trivial topology of the optical field. 

\subsection{Topological gaps}

The breaking of time-reversal symmetry opens gaps between the acoustic bands, bringing the vibrational excitations into a Chern insulator phase. 
This is confirmed by computing the topological invariant, known as the Chern number~\cite{Thouless1982}, $C_n = \frac{i}{2\pi}\int_{\rm BZ} d^2\vec k [\langle\partial_{k_x}m_{\vec k,n}|\partial_{k_y}m_{\vec k,n}\rangle-\langle\partial_{k_y}m_{\vec k,n}|\partial_{k_x}m_{\vec k,n}\rangle]$ for each acoustic band $n$ with energy eigenstates $\ket{m_{\vec k,n}}$. Here the integral is performed over the first Brillouin zone. For the two lowest-energy mechanical bands, one finds $C_1 = 1$ and $C_2 = 0$ [cf.~Fig.~\ref{Fig:DispRel} (b)].

As shown in Fig.~\ref{Fig:DispRel} (b), for weak OM interactions the gaps open at the symmetry points ${\bf K}$ and ${\bf K'}$, while for larger couplings, competing processes taking place with quasi momentum near the high-symmetry points ${\bf M}_1 = \frac{\pi}{2a}(1, -\frac{1}{\sqrt3})$, ${\bf M}_2 =- \frac{\pi}{2a}(1, \frac{1}{\sqrt3})$ and ${\bf M}_3= \frac{\pi}{a}(0, -\frac{1}{\sqrt3})$ close the gap again.
The dominant OM processes allowed by the conservation of angular momentum can be captured using a simple analytic model from which one accurately predicts the band gap
\begin{equation}
	\epsilon ={\rm Min}\left[ \frac{\delta_{\rm OM}}{2}\left(\sqrt{1+\frac{4G^2}{\delta_{\rm OM}^2}}-1\right), K\right],
	\label{Eq:Gap}
\end{equation}
for $G<G_c$, where
\begin{equation} \label{Eq:Gc}
	G_c/K = \sqrt{3\delta_{\rm OM}/2K},
\end{equation} 
is the critical coupling above which the gap decreases (cf.~Appendix~\ref{App:Gap}). 
Here $\delta_{\rm OM} = -\Delta - 2J - \omega_M - K$ is the detuning between the lowest optical band and the mechanical Dirac points in the noninteracting limit [cf.~Fig.~\ref{Fig:DispRel} (b)].

From the above expressions it follows that the band gap $\epsilon$ reaches the maximum value  $\epsilon=K$ for a detuning $\delta_{\rm OM}$ above the threshold $\delta^{\rm th}_{\rm OM}=2K$ and driving strengths in a finite range $G_{\rm min}\leq G\leq G_c$ where $G_{\rm min}=\sqrt{K^2+\delta_{\rm OM}K}$.
In Fig.~\ref{Fig:TopGap} (a), we show $\epsilon$ as a function of $\delta_{\rm OM}$ and $G$ while explicitly indicating $G_c$ and $G_{\rm min}$. In panel (b), we show $\epsilon$ as a function of $G$ for $\delta_{\rm OM}/K = 1.5, 2.0$.
For experimentally relevant phonon hopping rates $K/2\pi\sim 100$ MHz, the minimal coupling $G_{\rm min} \approx \sqrt{3}K$ at threshold $\delta^{\rm th}_{\rm OM}$ is reached with a number of intra-cavity photons $n_c \sim 0.75 \times 10^6$. While generally challenging, we note that recent experiments suggest that diamond structures~\cite{Burek2016} are more compliant to stronger drives compared to silicon-based systems~\cite{Safavi2010, Gavartin2011, Safavi2014, Kalaee2016}.

\subsection{Edge channels}
For a finite size system, the existence of separated energy bands with non-trivial Chern number is associated with a set of topologically protected chiral edge states, which propagate along the boundaries of the OM array. Specifically, the difference between the number of such edge states propagating clockwise and anti-clockwise is given by the sum over the topological invariant $C_n$ associated with all lower-energy bands.

To study in more detail the properties of these edge channels in the present setup, we consider in Fig.~\ref{Fig:EdgeState} (a) a stripe of infinite length along $x$ with straight edges at $y=0$ and $y = -(N_y-1) \sqrt3 a$. Here $N_y$ is the number of unit cells along $\vec R_1$ and the upper straight boundary is obtained by excluding the cavities $A$ of all cells at $y=0$. 
For this geometry, the full OM crystal Hamiltonian $\hat H_{\rm OMC}$ in Eq.~\eqref{Eq:HOMC} can be diagonalized within each quasi-momentum $k_x$ subspace, allowing us to capture the dispersion relation and the structure of the edge states localized at the boundaries.
Details of the diagonalization are  presented in Appendix~\ref{App:Hk}.
In Fig.~\ref{Fig:EdgeState} (b), we show an example of the mechanical band structure as a function of $k_x$ for $G=2K$, 
$\delta_{\rm OM} = 3K$ and $\Delta\theta = 3\pi/2$. 
On the upper boundary, a single edge state is present and its dispersion relation $\omega_E(k_x)$ is shown by the black curve crossing the gap for $\pi/2 < k_xa \leq \pi$. 
The group velocity of phonons propagating along this channel is given by
\begin{equation}
	v_g(k_x) = \frac{\partial \omega_{E}(k_x) }{\partial k_x},
\end{equation}
and is positive for the phase pattern $\Delta\theta = 3\pi/2$.
The situation is completely symmetric for the edge state at the lower boundary: its energy crosses the gap for $0 < k_xa \leq \pi/2$ and it has a negative group velocity.

For the purpose of using such edge modes as phononic quantum channels, two other key properties must be taken into account: their penetration depth into the bulk and how strongly they are hybridized with the optical bands. The latter plays an important role for dissipation  and the former characterizes how strongly the edge modes couple to the TLSs. In general, we can write the annihilation operator for an edge-state excitation with quasi-momentum $k_x$ as
\begin{align} \label{Eq:bEdge}
	\hat b_{\rm E}(k_x) = \sum_{s,m} &e^{-\frac{ma}{\xi(k_x)}-i\phi_m(k_x)}\\
	&\times [  u_{s}(k_x) \hat b_{s,m}(k_x) + v_{s}(k_x) \hat a_{s,m}(k_x)], \nonumber
\end{align}
where $\hat b_{s,m}(k_x)$ [$\hat a_{s,m}(k_x)$] is the phononic (photonic) annihilation operator acting on the basis $s = \{ A,B,C \}$ of the $m^{\rm th}$ unit cell along $\vec R_1$ with quasi-momentum $k_x$. The upper boundary corresponds to $m=0$.
The coefficients $u_s(k_x)$ and $v_s(k_x)$ are the respective mechanical and optical probability amplitudes, $\xi({k_x})$ is the penetration depth and $\phi_m(k_x)$ is a generic phase. 
We define the optical fraction of the edge state as
\begin{equation}
	P_{\rm opt}(k_x) = \sum_{s} \frac{|v_{s}(k_x)|^2}{1 - e^{-\frac{2a}{\xi(k_x)}}} \leq 1,
\end{equation}
where the upper bound is set by the normalization condition. 

In the view of a state transfer between TLS embedded in the outermost cavities along the boundaries of the array, we are mostly interested in the edge-state excitation that lies within the gap and has the smallest penetration depth and photonic amplitude. This condition motivates us to define $k_0$ such that $u_{s_0}(k_0)$ is maximal [cf.~Fig.~\ref{Fig:EdgeState} (e)], where $s_0$ represents the outermost cavities along the boundary.
In Fig.~\ref{Fig:EdgeState}, we show the optical fraction, the penetration depth $\xi(k_x)$ and the group velocity, all evaluated for $k_x = k_0$. 
We finally note that for the straight edges considered, i.e.~where only the cavities $B$ and $C$ form the outermost layer of the boundaries, no $A$ cavities throughout the crystal supports an edge mode (i.e.~$u_A = 0$).

We conclude this section by noting several important differences between the results presented here and those in Ref.~\cite{Peano2015}.
In this work, we consider positive hopping amplitudes $K>0$ and $J>0$ which results in inverted dispersion relations compared to Ref.~\cite{Peano2015}. As a consequence, the  lowest-energy optical band is flat (for $G=0$). This feature changes qualitatively  the  OM interaction. In particular,  the band gap $\epsilon$ and the optical fraction $P_{\rm opt}$ become independent of $J$  for large $J/\delta_{\rm OM}$. In contrast, these quantities are  suppressed as $\sqrt{K/J}$ and $K/J$, respectively, for negative hopping amplitudes~\cite{Peano2015}. Since $J/K\sim10^4$ is of the order of the ratio between the speed of light and sound in the material, this allows us to reach much larger band gaps at the expense of larger optical fraction.  Finally, for positive hopping amplitudes, the driven optical mode 
coincides with the lowest-energy band, such that the detuning from the drive frequency is considerably smaller than for the case of the highest-energy band considered in Ref.~\cite{Peano2015}.  Due to this  reduction of the detuning by about $6J\sim 100$ GHz, the necessary power of the external drive to reach $n_c \sim 10^6$ is strongly reduced.

	
\begin{figure}[t]
	\begin{center}
	\includegraphics[width= 0.95\columnwidth]{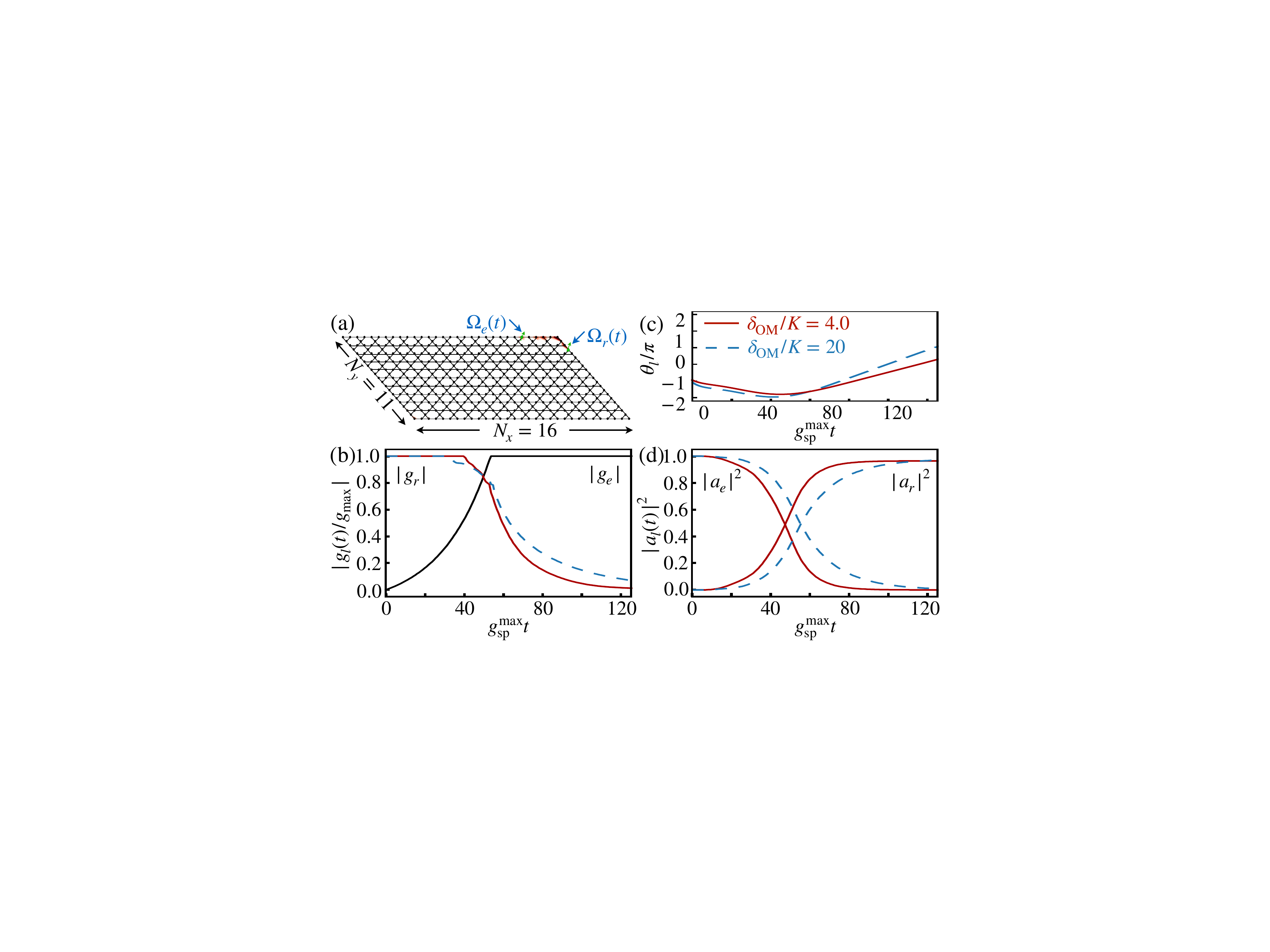}
	\end{center}
	\vspace{-0.5cm} 
	\caption {State transfer protocol. (a) Schematic of the state transfer over eight cavities around a corner for two scenarios: (1) $G=2K$, $\delta_{\rm OM} = 4K$ and $Q_C=5\times 10^7$ (solid lines) and  (2) $G=2K$, $\delta_{\rm OM} = 20K$ and $Q_C=10^7$  (dashed lines). (b) Time dependent coupling rates for the emitting and receiving defects. The emitting pulse is identical in both scenarios (black solid line). (c) The phase of the receiving pulses. (d) Amplitudes of the TLS as a function of time. 
	In all figures, we have considered $\omega_M / K = 460$, $\omega_C / K = 2\times10^{6}$, $J/K = 200$ and $g^{\rm max}_{\rm sp}/K = 0.06$.
}
\label{Fig:StateTransfer}
\end{figure}


\section{Quantum state transfer}

So far, we have focused solely on the OM cavities without considering finite couplings to the TLS. 
In this section, we exploit the time-dependent spin-phonon coupling $g_{\rm sp}(t)$ and the acoustic chiral edge states to transfer an arbitrary quantum state between any pairs of TLS embedded in distant cavities along the boundaries of the structure.

In this protocol, only the emitting ($e$) and receiving ($r$) defects are driven, i.e.~only $g_{\rm sp}^{(l)}(t) \neq 0$ with $l = \{e,r\}$, while all the other undriven centers are far off-resonance with any mechanical excitations. We also consider a low-temperatures environment $T \leq \hbar\omega_M/k_B$ (corresponding to $T \lesssim 1$K for SiV centers) in which case the system remains in the single excitation subspace.
Finally, we account for dissipative processes by including photon and phonon loss in every cavities of the crystal. We denote by $\kappa_C = \omega_C/Q_C$ ($\kappa_M = \omega_M/Q_M$) the photon (phonon) decay rate where $Q_C$ ($Q_M$) is the optical (mechanical) quality factor. 
By restricting the dynamics to the single-excitation subspace, we can account for losses by simply considering a non-unitary time evolution by substituting $-\Delta\rightarrow -\Delta - i\kappa_C/2$ ($\omega_M \rightarrow \omega_M - i\kappa_M/2$) in Eq.~\eqref{Eq:Hnetwork}.

\subsection{Markovian channel}
In the limit of weak spin-phonon couplings [$g^{(l)}_{\rm sp}(t) \ll K$], the topological phase of sound described in the previous section is approximately unperturbed by the TLS. 
Moreover, for resonance frequencies $\omega_0$ of both TLS deep within the topological band gap, the defects only couple efficiently to the acoustic edge modes.
In this limit, the coherent dynamics of the state transfer protocol can be described by the effective Hamiltonian
\begin{align} \label{Eq:Heff}
& \hat H_{\rm st}(t) = \sum_k \omega_E(k) \hat b_E^\dag (k) \hat b_E(k) + \sum_{{l = e,r}}\frac{\omega_0}{2}\hat \sigma_z^{(l)}\\
	& + \frac{1}{\sqrt N}\sum_k\sum_{l = e,r} \left[ g_{\rm eff}^{(l)}(t)\hat \sigma_+^{(l)}b_{E}(k)e^{2ikn_la} + {\rm H.c.} \right], \nonumber 
\end{align}
where $k$ and $n_l$ represent the quasi-momentum of the edge state and the position of the TLS $l$ along the edge of $N$ cavities, respectively. 
The effective spin-phonon coupling $g_{\rm eff}^{(l)}(t) = u_{s_l}(k)e^{-m_la/\xi(k)-i\phi_{m_l}(k)}g_{\rm sp}^{(l)}(t)$ depends on the distance of the defect from the boundary ($m_l$) and captures the properties of the edge modes previously derived in the context of the semi-infinite stripe.
Although the structure supporting the state transfer is a finite 2D crystal [cf.~Fig.~\ref{Fig:StateTransfer} (a)], Eq.~\eqref{Eq:Heff} is valid for defects positioned far from any dislocations, e.g., a corner. 
Similarly, one can estimate the decay rate of the chiral channel as $\kappa_E(k) = |u_s|^2\kappa_M/[1 - e^{-2a/\xi(k)}] + P_{\rm opt}(k)\kappa_C$.
For the scenarios considered in this work, where the mechanical frequencies are in the GHz regime while the optical modes are in the hundreds of THz, $\kappa_C/\kappa_M \sim \omega_C/\omega_M \sim 10^{4}$ in cases of similar quality factors. As a consequence, the optically induced decay rate is expected to exceed by far the intrinsic mechanical loss. 

Considering the single-excitation ansatz
\begin{align}
	\ket{\psi(t)} & = \alpha \ket{0} + \beta e^{-i\omega_0t}[ a_e(t)\hat \sigma_{+}^{(e)} + a_r(t)\hat \sigma_{+}^{(r)} \\
	& \qquad + \sum_k a_k(t) \hat b_E^\dag(k)]\ket{0}, \nonumber
\end{align}
where $\ket0$ represents the vacuum state with both centers being in their lowest-energy state and no acoustic excitations in the waveguide, a perfect transfer of an arbitrary state corresponds to $a_e(t_0) = a_r(t_f) = 1$ and $a_e(t_f) = a_r(t_0) = 0$. Here $t_0$ and $t_f$ are the initial and final times of the protocol, respectively.
In the case where the TLS see a constant density of states of the acoustic modes, it is possible the apply the standard Born-Markov approximation to the Schr\"odinger equation $i\hbar\partial_t \ket{\psi(t)} = \hat H_{\rm ss}(t)\ket{\psi(t)}$ (cf.~Appendix~\ref{App:1DWG}), leading to the following local equations of motion 
\begin{equation} \label{Eq:EOM}
	\partial_t  a_l(t) = -\frac{\gamma_l(t)}{2}a_l(t) - \sqrt{\gamma_l(t)}e^{i\theta_l(t)}f_{{\rm in}, l}(t).
\end{equation}
Here, the transfer rate between the chiral waveguide and the defects is
\begin{equation} \label{Eq:Gamma}
	\gamma_l(t) = 2\frac{|g_{\rm eff}^{(l)}(t)|^2}{v_g/a},
\end{equation}
with $v_g \equiv v_g(k_0)$ and $\theta_l(t) = \arg[g_{\rm eff}^{(l)}(t)]$. The input field of the receiving node is $f_{\rm in, r}(t) = f_{\rm out, e}(t - \tau_p)e^{i\phi_p}$ with $\tau_p$ and $\phi_p$ the time and phase acquired during the propagation from the emitter to the receiver. In the case of a perfectly straight edge, $\tau_p = 2(n_r - n_e)a/v_g$ and $\phi_p = k_02(n_r - n_e)a$. 
The strategy to achieve a high-fidelity state transfer is to control in time $g_{\rm eff}^{(r)}(t)$ in order to suppress the output field $f_{{\rm out}, r}(t)  = f_{{\rm in}, r}(t) +  \sqrt{\gamma_r(t)}e^{-i\theta_r(t)}a_r(t)$ of the receiver. Thus in this idealized limit, the state transfer-problem becomes equivalent to the scenario discussed in the original work by Cirac {\it et al.}~\cite{Cirac1997} and similar optimized pulse shapes can be used to achieve close-to-unity state transfer fidelities. The main limitation then arises from propagation losses and the ratio between the TLS decoherence rate and the maximal transfer rate $\gamma_{\rm max}$ that one can reach in a specific implementation.

	
\begin{figure}[t]
	\begin{center}
	\includegraphics[width= 0.70\columnwidth]{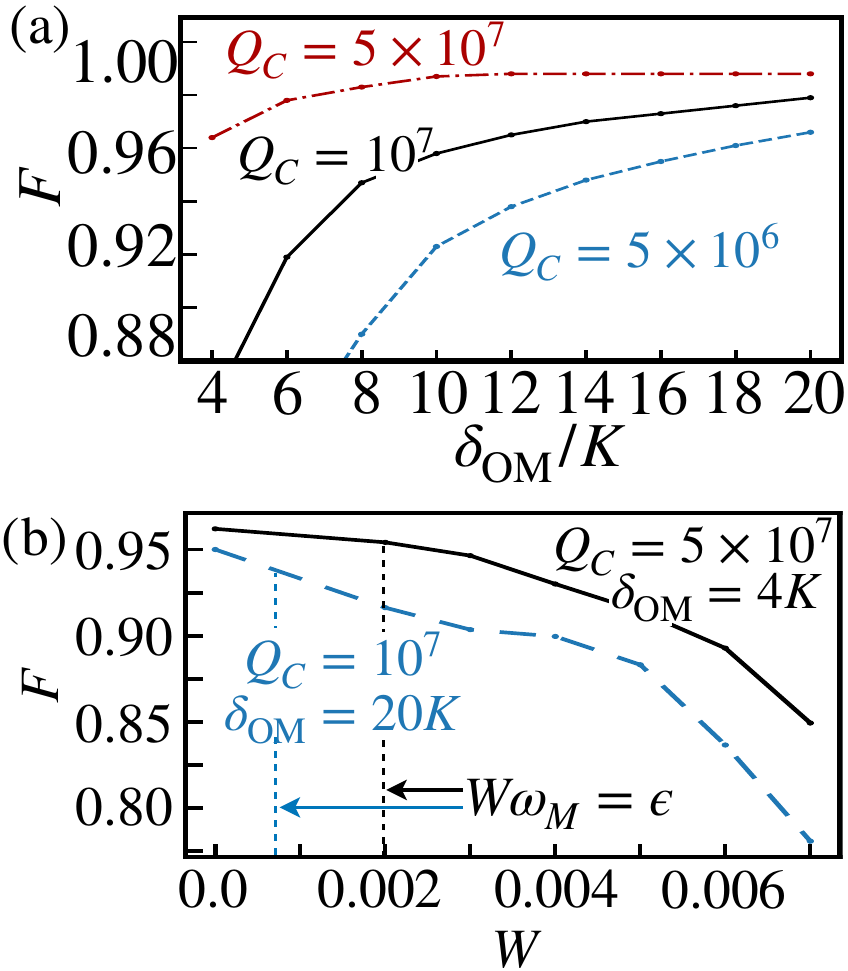}
	\end{center}
	\vspace{-0.5cm} 
	\caption {State transfer fidelity. (a) State transfer fidelity $F = |a_r(t_f)|^2$ as a function of $\delta_{\rm OM}$ for optical quality factors $Q_C = 5.0\times10^6, 10^7, 5\times 10^7$. Here, $G=2K$ and $Q_M = 10^6$. (b) Fidelity as a function of the disorder strength $W$ for $G=2K$, $\delta_{\rm OM} = 4K$ ($\delta_{\rm OM} = 20K$) and $Q_C= 5\times 10^7$ ($Q_C= 10^7$) plotted in solid line (dashed).
	The vertical lines represent the disorder strengths for which $W\omega_M = \epsilon$ where $\epsilon/K = 1.0$ for $\delta_{\rm OM}/K = 4$ and $\epsilon/K = 0.34$ for $\delta_{\rm OM}/K = 20$.
	In all figures, we have considered $\omega_M / K = 460$, $\omega_C / K = 2\times10^{6}$, $J/K = 200$ and $g^{\rm max}_{\rm sp}/K = 0.02$.}
\label{Fig:Fidelity}
\end{figure}


\subsection{Exact evolution}
While the above description properly highlights the physics underlying the state-transfer protocol, it becomes exact only for extremely weak couplings to perfect Markovian 1D channels. 
In contrast, we here perform no approximations and numerically simulate the full dynamics of the time evolution as governed by Eq.~\eqref{Eq:Hnetwork}. 
We use a slowly varying pulse for the emitter $g^{(e)}_{\rm sp}(t)/g^{\rm max}_{\rm sp} = \min[1, e^{(t - 4) / 2}]$ with a weak maximal coupling $g^{\rm max}_{\rm sp}/K = 0.06$. 
The optimal pulse for the receiver $g_{\rm sp}^{(r)}(t)$ is then determined by maximizing at every time step the amplitude of the receiver $|a_r(t)|$.
In Fig.~\ref{Fig:StateTransfer} (b)-(d), we show two examples of state transfers over a distance of eight cavities along the edge of a crystal with $N_x = 16$ ($N_y = 11$) unit cells along $x$ ($y$). Specifically, in both examples the emitter and the receiver are located on different edges of the crystal [cf. Fig.~\ref{Fig:StateTransfer} (a)], such that the non-trivial transfer of excitations around a corner is included in the simulations. 
We compare two scenarios: the first one with higher cavity quality factor $Q_C = 5.0 \times 10^7$ and strong OM coupling $G = 2K$ with $\delta_{\rm OM} = 4K$; and the second with lower $Q_C = 10^7$ and more detuned OM coupling $G = 2K$ and $\delta_{\rm OM} = 20K$.
In the first scenario, the edge state is more localized and has a slower group velocity (see Fig.~\ref{Fig:EdgeState}), leading to a faster state-transfer via a stronger $\gamma_{\rm max}/ g^{\rm max}_{\rm sp}\sim 0.03$ [cf.~Eq.~\eqref{Eq:Gamma}].
However, the larger optical hybridization and longer time spent in the waveguide make the optically-induced decay rate more detrimental, hence the need of higher $Q_C$.
In the second scenario, the transfer is slower with $\gamma_{\rm max}/ g^{\rm max}_{\rm sp}\sim 0.006$, but more resilient to dissipation.

In Fig.~\ref{Fig:Fidelity} (a), we analyze the maximal fidelity $F = |a_r(t_f)|^2$ as a function of the detuning $\delta_{\rm OM}$ for $G = 2K$ for $Q_C = \{ 0.5, 1.0, 5.0\}\times 10^7$. 
It highlights the larger optically-induced decay rate for smaller detunings. In the short-time limit, one can approximate the optical loss as $1-F \sim N_s P_{\rm opt}\kappa_C a/ v_g$ with $N_s$ the number of traveled cavities.
For larger detunings, the optical loss is reduced, but the smaller decay rates require longer time $t_f$ to transfer the state, which can become an issue compared to the coherence time of the TLS.
As an example, for $K/2\pi = 100$ MHz and $g^{\rm max}_{\rm sp}/K = 0.06$, $t_f\sim 2\mu$s for $G=2K$ and $\delta_{\rm OM} = 4K$. This is still fast compared to the expected inhomogeneous dephasing times $T_2^*\sim 10-100\,\mu$s and much shorter than the intrinsic coherence time of $T_2\sim10$ ms demonstrated for SiV centers at low temperatures. 

\subsection{Disorder}

So far, we have consider identical parameters over the entire system, i.e. perfectly matched mechanical frequencies, detunings and OM couplings. In experiments, reaching a high level of homogeneity is challenging and any realizations is expected to have a certain level of disorder. 
We here analyze the robustness of the state transfer in presence of such imperfections within the system.
To do so, we consider all parameters to be slightly different for every cavities. For example, we consider $\omega_M^{(j)} = (1 + p_j)\omega$ where $p_j$ is randomly chosen from a uniform distribution ranging from $-W/2$ to $W/2$. The same applies for $\Delta$, $G$, $\kappa_C$ and $\kappa_M$. 
In Fig.~\ref{Fig:Fidelity} (b), we plot the state transfer fidelity as a function of $W$ averaged over 50 realizations of disorders. We compare the robustness for $G=2K$ and $\delta_{\rm OM} = 4K$, where the gap is $\epsilon = K$, to the scenario with $G=2K$ and $\delta_{\rm OM} = 20K$, where the gap is $\epsilon \approx 0.34K$. 
The state-transfer fidelity starts to decrease for disorder strengths large enough to close the topological gap, which is roughly set by $W\omega_M \gtrsim \epsilon$ [as indicated by the vertical lines in Fig.~\ref{Fig:Fidelity} (b)].
An additional advantage of working with larger OM interactions is thus the increased resilience to disorder due to the larger gap.

\section{Conclusion and outlook}

In this work, we have proposed and analyzed a 2D hybrid system where the acoustic excitations within a Kagome lattice of coupled OM cavities interact with spin degrees of freedom of point defects.
In this context, we have described the emergence of a topological phase where time-reversal symmetry is effectively broken for the  acoustic excitations as a result of the interplay between the OM interaction and the inter-cavity hopping. As a potential  application, we have shown that the resulting acoustic chiral edge states can serve as phononic quantum channels, which are purely unidirectional and robust with respect to onsite-disorder. Our analysis revealed how the key properties of such topological channels depend on the relevant OM coupling and detuning parameters and how an optimized trade-off between optical losses, propagation speed and disorder protection can be achieved. Apart from the considered example of SiV defects in diamond OM crystals, most of these considerations will be relevant as well for other types of qubits or other artificial realizations of topological systems. 

Beyond quantum communication applications, the proposed hybrid system provides a versatile platform to study interacting quantum many-body systems, where topological phases with broken time-reversal symmetry are combined with strong nonlinearities provided by the spin qubits. The rich physics expected for such interacting topological insulators is still little understood and could be probed in such setting in various parameter regimes and employing only static spin-phonon interaction, which are in general much simpler to engineer.

\section{Acknowledgements}

M.-A. L. thanks Dieter Jaksch for fruitful discussions. This research is supported by the National Research Foundation, Prime Minister’s Office, Singapore and the Ministry of Education, Singapore under the Research Centres of Excellence programme. It was also partially funded by Polisimulator project co-financed by Greece and the EU Regional Development Fund, the European Research Council under the European Union’s Seventh Framework Programme (FP7/2007–2013).

\appendix

\section{Rotating wave approximation and OM instability} \label{App:RWA}

Throughout this work we have neglected the effects of the OM off-resonant parametric type terms
\begin{equation}
\sum_j Ge^{-i\theta_j}\hat{a}_j\hat{b}_j+ {\rm H.c.}
\end{equation}
Such terms describe the creation and annihilation of photon-phonon pairs and have been dropped  during the derivation of Eq.~\eqref{Eq:HOMC} based on a rotating wave approximation. The processes that dominantly contribute to the corrections to the rotating wave approximation describe the creation (annihilation) of a phonon accompanied by the creation (annihilation) of a photon in the flat Kagome band. The typical oscillation frequency $\delta_K$ of these terms in the rotating frame is set by the distance of the flat optical band from its blue side band, $\delta_K\sim\delta_{\rm OM}+2\omega_M$. As a consequence the leading order corrections to the RWA  are of the order $\sim G/(\delta_{\rm OM}+2\omega_M)$. It is important to keep in mind that contrary to the usual case of time-reversal preserving OM systems such terms can lead to an OM instability even in the regime where they represent a small perturbation and when the driving is red detuned compared to all optical resonances,  see Ref. \cite{Peano2015} for a detailed analysis. The reason for this behavior is that the same optical mode couples to different mechanical modes on its blue and red sidebands. As a consequence the mechanical modes that couple only to the blue sideband of the flat Kagome band but not to its red sideband are subject to a small overall optical induced amplification with rate $\sim\kappa_C G^2/\delta_K^2$. This implies that a small mechanical decay rate of the order $\kappa_M\sim\kappa_C G^2/(\delta_K)^2$ is required to stabilize the system,
which is present for all parameter regimes considered in this manuscript.

\section{Negatively charged Silicon-Vacancy centers in Diamond}
\label{App:SiV}

In this section we describe in more detail the negatively-charged Silicon-vacancy center in diamond. More precisely, we focus on its electronic ground-state and its strain coupling to vibrational modes of its host crystal.
 
The molecular structure of the SiV center belongs to the $D_{3d}$ point group symmetry and their electronic ground state are formed by an unpaired hole of spin $S=1/2$ subjected to a strong spin-orbit interaction.
The resulting fourfold ground state subspace is comprised of two doublets, $\{\ket1\simeq  \ket{e_-,\downarrow}, \ket2\simeq |e_+,\uparrow\rangle\}$ and $\{\ket3\simeq|e_+,\downarrow\rangle, \ket4\simeq|e_-,\uparrow\rangle\}$, which are separated by $\Delta_{\rm SiV}/2\pi \simeq 46$ GHz~\cite{Hepp2014, HeppThesis}. Here, $\ket{e_{\pm}}$ are eigenstates of the orbital quasi-angular momentum operator $\hat L_z$ associated to a $2\pi/3$ rotation about the symmetry axis of the defect (taken to be along $z$), i.e.~$\hat R_{2\pi/3}\ket{e_{\pm}} = e^{-i\frac{2\pi}{3}\frac{\hat L_z}{\hbar}}\ket{e_{\pm}} = e^{\mp i\frac{2\pi}{3\hbar}}\ket{e_{\pm}}$. 
In the presence of a magnetic field $\vec B=B_0\vec e_z$, the Hamiltonian for a single SiV center is $(\hbar=1)$
\begin{equation}\label{Eq:HSiV}
\begin{split}
\hat H_{\rm SiV}  &= \omega_B  \ket2 \bra2 + \Delta_{\rm SiV}\ket3 \bra3 + (\Delta_{\rm SiV} + \omega_B)\ket4 \bra4\\
	&+\frac{1}{2}\left[\Omega(t)e^{i[\omega_{d} t + \phi(t)]} \left(\ket2 \bra3 +  \ket1 \bra4  \right) +{\rm H.c.}\right],
	\end{split}
\end{equation}
where $\omega_B=\gamma_sB_0$  and  $\gamma_s$ is the spin gyromagnetic ratio. 
In Eq.~\eqref{Eq:HSiV}, we have included a time-dependent driving field of frequency $\omega_d$ with a tunable Rabi-frequency $\Omega(t)$ and phase $\phi(t)$, which couples the lower and upper states of opposite spin. This drive can be implemented locally on individual defects either directly with a microwave field of frequency $\omega_{d}\sim\Delta_{\rm SiV}$~\cite{Pla2012}, or indirectly via an equivalent optical Raman process~\cite{Lemonde2018}. 

\subsection{Strain coupling to mechanical modes}
Within an OM cavity, we consider a single mechanical mode associated with a displacement profile $\vec u(\vec r)$.
In addition to modifying the indice of refraction seen by the optical mode, such deformation of the cavity modifies the electronic environment seen by the SiV center, resulting in the coupling of its orbital states $\ket{e_\pm}$~\cite{Sohn2017,Jahnke2015, Kepesidis2016}. The SiV-phonon coupling can be described by ($\hbar = 1$)
\begin{equation}
\hat H_{\rm int} = g_s \hat J_+ \hat b + {\rm H.c.},
\end{equation}
where $\hat J_- = (\hat J_+)^\dag = \ket1\bra3 + \ket2\bra4$ is the spin-conserving lowering operator and $g_s$ is the strain-induced coupling rate which is proportional to the local strain tensor $\epsilon^{ab}(\vec r) = \frac{1}{2}[ \frac{\partial }{\partial x_b} u_a(\vec r) + \frac{\partial }{\partial x_a}  u_b(\vec r)]$. 
The resulting coupling rate can be written as $g_s = 2\pi d \frac{x_{\rm ZPF}}{\lambda_{\Delta_{\rm SiV}}}\xi(\vec r_{\rm SiV})$, where $d/2\pi\sim 1$ PHz is the strain sensitivity~\cite{Jahnke2015, Sohn2017}, $x_{\rm ZPF} \sim 1$ fm is the mechanical zero point motion~\cite{Safavi2010T}, $\lambda_{\Delta_{\rm SiV}} \sim 200 $ nm the characteristic phonon wavelength in diamond and $\xi(\vec r_{\rm SiV})$ is the dimensionless strain distribution evaluated at the position of the SiV center $\vec r_{\rm SiV}$. 
From deformations $\vec u (\vec r)$ observed in previous experiments~\cite{Safavi2010} and state-of-the-art positioning of SiV defects~\cite{Schroder2016}, we expect $\xi(\vec r_{\rm SiV}) \sim 1$, leading to interaction rates as large as $g_s/2\pi \sim 30$ MHz. This estimation is consistent with finite-element simulations performed for 1D diamond nano-cavities~\cite{Lemonde2018}.
For matching frequencies ($\Delta_{\rm SiV} = \omega_M$) and mechanical quality factors $Q\sim10^5$, strong-coupling regime $g_s > \omega_M/Q, 1/T_2^*$ should be reached, allowing coherent excitation transfer between the SiV centre and the mechanical resonator.

\subsection{Time-dependent effective spin-phonon coupling}
In the specific case of a state transfer protocol, one has to control in time an effective coupling between the spin degree of freedom, encoded in the two lowest-energy ground-states $\ket1$ and $\ket 2$, and the mechanical mode. This can be performed by an off-resonant driving of the state $\ket3$ [cf.~Fig.~\ref{Fig:Schema} (b) and Eq.~\eqref{Eq:HSiV}], leading to a standard three-level $\Lambda$ atomic system. For large drive detunings $\delta = \omega_0 - \Delta_{\rm SiV}$, i.e.~$|\delta| \gg |g_s|, |\omega_M-\omega_0|, |\Omega|$ with $\omega_0 = \omega_d + \omega_B$ the frequency of the emitted phonons, the higher-energy state $\ket3$ can be adiabatically eliminated resulting in an effective time-dependent spin-phonon coupling $g_{\rm sp}(t) = g_s\Omega(t)/\delta$.
Assuming $0\leq \Omega(t)/2\pi < 100$ MHz and $\delta/2\pi \lesssim 400$ MHz, this rate can be tuned between $g_{\rm sp}=0$ and a maximal value of $g_{\rm sp}^{\rm max}/2\pi\sim 7$ MHz, which is still large enough to reach the strong coupling regime.

\section{Diagonalization of the OM crystal Hamiltonian in the momentum space} \label{App:Hk}

In this appendix, we provide details of the diagonalisation in the quasi-momentum space of the OM Hamiltonian, $\hat H_{\rm OMC}$ introduced in Eq.~\eqref{Eq:HOMC}, in the case of an infinite 2D array and a semi infinite stripe.
Focusing on the Kagome lattice architecture, the sum over all sites $j$ of the crystal in Eq.~\eqref{Eq:HOMC} can be expanded into the sum over all unit cells of the triangular lattice and the three basis cavities within each cells, i.e.~$j \rightarrow \{j, s\}$ corresponding to the cavity $s = \{ A, B, C \}$ in the unit cell centered at $\vec R_j = m_j \vec R_1 + n_j\vec R_2$.  Doing so, $\hat H_{\rm OMC}$ reads
\begin{align}
	& \hat H_{\rm OMC} = \sum_{j,s} ( \omega_M \hat b_{s,j}^\dag \hat b_{s,j} - \Delta \hat a_{s,j}^\dag \hat a_{s,j} \!+\! G e^{i\theta_{s}} \hat a_{s,j}^\dag \hat b_{s,j}  \\
	& \quad + Ge^{-i\theta_s} \hat b_{s,j}^\dag \hat a_{s,j} )
	+\sum_{n.n.} (K \hat b_{r,i}^\dag \hat b_{s,j} + J \hat a_{r,i}^\dag \hat a_{s,j} + {\rm H.c.} ). \nonumber
\end{align}
Here, $\sum_{n.n.}$ represents the sum over the nearest neighbors.

\subsection{Infinite 2D crystal}

In the limit where the system is an infinite 2D array with $N \rightarrow \infty$ cavities, one can define 
\begin{equation} \label{Eq:bk2D}
	\hat b_s(\vec k) = \frac{1}{\sqrt N} \sum_{j} e^{-i\vec k \cdot \vec R_j} \hat b_{s,j},
\end{equation}
which destroys an excitation with the conserved quasi-momentum $\vec k$. The same definition applies for the optical modes, leading to
\begin{equation}
	\hat H_{\rm OMC} = \sum_{\vec k}\hat H(\vec k) = \sum_{\vec k} \hat H_M(\vec k) + \hat H_C(\vec k) + \hat H_G(\vec k),
\end{equation} 
where the mechanical Hamiltonian
\begin{align}
	\hat H_M(\vec k) = & \sum_s \omega_M \hat b^\dag_{s}(\vec k)\hat b_s(\vec k)  + \left\lbrace K(1 + e^{-i \vec k \cdot \vec R_1}) \hat b^\dag_A(\vec k)\hat b_B(\vec k) \right.\nonumber \\
	& + K(1 + e^{-i \vec k \cdot [\vec R_1 + \vec R_2]}) \hat b^\dag_A(\vec k)\hat b_C(\vec k) \\
	& \left. + K(1 + e^{-i \vec k \cdot \vec R_2}) \hat b^\dag_B(\vec k)\hat b_C(\vec k) + {\rm H.c.} \right\rbrace\nonumber .
\end{align}
Here, $\hat b^\dag_s(\vec k) =[ \hat b_s(\vec k)]^\dag$ and the equivalent form applies to the photons $\hat H_C(\vec k)$. The OM interactions read
\begin{align}
	\hat H_G(\vec k) = G\sum_s  \hat b^\dag_s(\vec k)\hat a_s(\vec k)e^{i\theta_s} + {\rm H.c.}
\end{align} 
Since $H_{\rm OMC}$ is quadratic, one can fully solve the excitation spectrum considering only a single excitation for which $\hat H(\vec k)$ is a $6\times6$ matrix.  Diagonalising $\hat H(\vec k)$ for every $\vec k$ within the first Brillouin zone of the Kagome lattice leads to a six-band dispersion relation as shown in Fig.~\ref{Fig:DispRel} (a).

\subsection{Semi-infinite stripe}

For a stripe that is infinite in the $x$ direction (along $\vec R_2$) with $N_y$ unit cells along $\vec R_1$, only the quasi momentum along $x$ ($k_x$) is conserved and the proper expansion for $\hat b_{s,j}$ (same for $\hat a_{s,j}$) reads
\begin{equation}
	\hat b_{s,j} = \frac{1}{\sqrt{N_x}} \sum_{k_x = -\pi/2a}^{\pi/2a} e^{i 2nk_xa}\hat b_{s,m_j}(k_x).
\end{equation}
Here $N_x \rightarrow \infty$ is the number of unit cells along $x$ and $\hat b_{s,m}(k_x)$ is the destruction operator for a mechanical mode with quasi-momentum $k_x$ in cavity $s$ of the $m^{\rm th}$ unit cell along $\vec R_1$. 
We thus consider a strip that goes from $y=0$ to $y = -N_y\sqrt3 a$ and increasing $m$ means to go along $-y$.

Similar to the infinite 2D case, one gets
\begin{equation}
	\hat H_{\rm OMC} = \sum_{k_x}\hat H(k_x) = \sum_{k_x} \hat H_M(k_x) + \hat H_C(k_x) + \hat H_G(k_x),
\end{equation}
with
\begin{align}
	& \hat H_M(k_x) = \sum_s\sum_{m=1}^{N_y} \omega_M \hat b^\dag_{s,m}(k_x)\hat b_{s,m}(k_x) \\
	& + K \sum_{m=1}^{N_y} \left\{ \hat b^\dag_{A,m}(k_x) [  \hat b_{B,m-1}(k_x)e^{2ik_xa} + \hat b_{B,m}(k_x) \right. \nonumber \\
	& \left. + \hat b_{C,m}(k_x) \!+\! \hat b_{C,m-1}(k_x)] + \hat b^\dag_{B,m}(k_x)\hat b_{C,m}(k_x) [1\!+\!e^{-2ik_xa} \right\} \nonumber \\
	& + \hat b^\dag_{B,0}(k_x) \hat b_{C,0}(k_x)[1+e^{-2ik_xa}] + {\rm H.c.} \nonumber 
\end{align}
The last line describes the hoppings within the first unit cell and so determines the form of the edge. In that case, the site $A$ is missing which leads to a straight edge as pictured in Fig.~\ref{Fig:EdgeState} (b). The optical Hamiltonian $\hat H_C(k_x)$ adopts the same form. 

Within the single excitation subspace, $H_{k_x}$ is a matrix of dimension $3N_y-1$ and its diagonalisation leads to the dispersion relation shown in Fig.~\ref{Fig:EdgeState} (a).
For finite $G$ and a phase pattern $\Delta\theta = \pm3\pi/2$, the edge states appear in the energy spectrum and can be expressed within the same basis, i.e.
\begin{equation}
	\hat b_{\rm E}(k_x) \!=\! \sum_{s,m} e^{-\frac{ma}{\xi(k_x)}} e^{-i\phi_m(k_x)}[ u_{s} \hat b_{s,m}(k_x)\! +\! v_{s} \hat a_{s,m}(k_x) ].
\end{equation}
Here, the coefficients $u_s$ and $v_s$ are the mechanical and optical probability amplitudes on the basis $s$, respectively. The edge state decays exponentially within the bulk with a penetration depth $\xi(k_x)$ and phases $\phi_m(k_x)$. 

\section{Topological gap} \label{App:Gap}

In this section, we derive in more detail the effective models to describe the opening and closing of the topological gap. We consider an infinite 2D array and utilize the modes derived in Eq.~\eqref{Eq:bk2D}. 

As described in the main text, at the high symmetry points ${\bf K}$ and ${\bf K'}$, the eigenstates of the Kagome lattice have also well-defined quasi-angular momentum (upon rotation of $2\pi/3$), i.e.~$\hat R_{2\pi/3}\ket{m_{{\bf K},\sigma}} = e^{-i\sigma \frac{2\pi}{3\hbar}}\ket{m_{{\bf K},\sigma}}$ with $\sigma = \{ -1,0,1 \}$ (same for ${\bf K}'$). In that case
\begin{equation}
	\hat b_{{\bf K},\sigma} = \sum_j  \frac{e^{-i{\bf K} \cdot \vec R_j}}{\sqrt{3N}} \left(\hat b_{A,j} + e^{-i\sigma2\pi/3} \hat b_{B,j} + e^{i\sigma2\pi/3} \hat b_{C,j} \right).
\end{equation}
In addition, a phase pattern of the drive $\Delta\theta = \theta_A-\theta_B = \theta_B - \theta_C = \theta_C-\theta_A = \pm2\pi/3$ means that only the optical mode $\ket{o_{{\bf \Gamma}, \pm}}$ is driven.

Following the conservation of the total angular momentum, only few OM processes are allowed. For example, given $\Delta\theta = 2\pi/3$, the lowest-energy optical eigenstate is $\ket{o_{{\bf K},+}}$ and can only interact with the mechanical eigenstate $\ket{o_{{\bf K},+}}$ at the Dirac point $\omega_M + K$, leading to a simple two-mode effective model
\begin{equation}
	H_{\rm eff, {\bf K}} = \delta_{\rm OM} a^{\dag}_{{\bf K}, -}a_{{\bf K}, -} + G(a^{\dag}_{{\bf K}, -}b_{{\bf K}, +} + a_{{\bf K}, -}b^\dag_{{\bf K}, +}).
\end{equation}
Here $a_{{\bf K}, -} $ destroys a photon in mode $\ket{o_{{\bf \Gamma}, \mp}}$ and the effective Hamiltonian is written in a frame that rotates at the frequency $\omega_M + K$. Each time a phonon is destroyed, a photon of quasi-angular momentum $\sigma = +$ is also absorbed. $H_{\rm eff, {\bf K}}$ is easily diagonalized and leads to the gap 
presented in Eq.~\eqref{Eq:Gap}. Note that  the highest mechanical band remains untouched with $E_{m,2}({\bf K}) = \omega_M + K$.

For larger coupling rates $G$, processes occurring near quasi-momentum ${\bf M}_1$, ${\bf M}_2$ and ${\bf M}_3$ start to play a role (in what follows we omit the subscript for clarity). Those high-symmetry points are invariant under ${\cal C}_2$ rotations. As a consequence the normal modes are divided into symmetric and anti-symmetric normal modes at these points. For this reason the anti-symmetric mechanical band $E_{m,2}({\bf M})$ do not interact with the symmetric  optical band $E_{o,1}({\bf M})$. The consequence is that no matter how large the gap at ${\bf K}$ and ${\bf K'}$ becomes, the middle mechanical band stays at $E_{m, {\bf M}} = \omega_M$ and thus bounds the total gap at $\epsilon_{\rm max} = K$.

Moreover, the allowed interaction between the optical band $E_{o,1}({\bf M})$ and the highest mechanical band $E_{m,3}({\bf M})$ has the net effect to push down the mechanical band which closes the gap. In order to accurately capture the processes, we also need to include the lowest mechanical band, leading to a 3-mode effective model
\begin{align}
	H_{\rm eff, {\bf M}} & = \delta_{\rm OM} a^{\dag}_{{\bf M}, 1}a_{{\bf M}, 1} - 3K b^{\dag}_{{\bf M}, 1}b_{{\bf M}, 1} \nonumber \\
	& + \left( \frac{G}{2} a^{\dag}_{{\bf M}, 1}b_{{\bf M}, 1} + \frac{\sqrt{3}G}{2} a^{\dag}_{{\bf M}, 1}b_{{\bf M}, 3} + {\rm H.c.} \right).
\end{align}
From $H_{\rm eff, {\bf M}}$, one can find the critical value $G_c$ at which the gap starts to close, leading to Eq.~\eqref{Eq:Gc}.

\section{State transfer in a 1D Markovian waveguide} \label{App:1DWG}

In this section, we write a simple model for two TLS weakly coupled to a 1D chiral channel within the Born-Markov approximation. Doing so, we derive the effective transfer rate between the emitting TLS and the waveguide as well as the dissipation rates due to photons and phonons loss. We connect the important results to the group velocity $v_g$, penetration depth $\xi$ and optical fraction of the edge state $P_{\rm opt}$, all shown in Fig.~\ref{Fig:EdgeState}. 

We consider both the emitting and receiving defects, denoted by the subscripts $\{e, r\}$ respectively, to be localized in the outermost unit cells along the edge, i.e.~$m=0$ in Eq.~\eqref{Eq:bEdge}. Only keeping the edge state from the OM array, we write the simplest Hamiltonian
\begin{equation}
	\hat H_{\rm 1D} = \hat H _{\rm TLS} +  \hat H_{\rm E} + \hat H_{\rm int}
\end{equation}
with
\begin{align}
	 \hat H _{\rm TLS} & = \frac{\omega_0}{2} (\sigma_{e,z} + \sigma_{r,z}), \nonumber \\
	 \hat H_{\rm E} & = \sum_{k_x} \omega_E(k_x) b^\dag_E(k_x)b_E(k_x), \\
	 \hat H_{\rm int} & = \frac{1}{\sqrt N_x}\sum_{k_x} \left[u_{s_e}(k_x) e^{2ik_xj_ea}g^{(e)}_{\rm sp}(t)\sigma_{e,+} b_E(k_x) \right. \nonumber \\
	   & \left. + u_{s_r}(k_x) e^{2ik_xj_ra}g^{(r)}_{\rm sp}(t)\sigma_{r,+} b_E(k_x) + {\rm H.c.} \right]. \nonumber
\end{align}
Here $j_e$ ($j_r$) and $s_e$ ($s_r$) indicate the unit-cell position along the edge and which basis the emitting (receiving) TLS is coupled to, with corresponding coupling strength $g_{\rm sp}^{(e)}$ ($g_{\rm sp}^{(r)}$).

We consider the single-excitation ansatz
\begin{align}
	&\ket{\psi(t)}  = \alpha \ket{0} \\
	& + \beta e^{-i\omega_0t}\left[ a_e(t)\sigma_{e,+} + a_r(t)\sigma_{r,+} + \sum_{k_x} a_{k_x}(t) b_E^\dag(k_x)  \right]\ket{0}, \nonumber 
\end{align}
where $\ket0$ represents the ground state of the whole system. The Schr\"odinger equation for the edge modes leads to
\begin{align}
	&a_{k_x}(t) = e^{-i(\omega_E(k_x) - \omega_0)(t-t_0)} a_{k_x}(t_0) \\
	&	\qquad - i\frac{1}{\sqrt N_x} \int_{t_0}^t d\tau e^{-i(\omega_E(k_x) - \omega_0)(t-\tau)} \nonumber \\
	& \left[ u^*_{s_e}(k_x) e^{-2ikj_ea}g_{\rm sp}^{(e)*}(\tau)a_e(\tau) + u^*_{s_r}(k_x)e^{-2ikj_ra}g_{\rm sp}^{(r)*}(\tau)a_r(\tau )\right]. \nonumber
\end{align}
Using this result in the equation for the receiver's cavity and performing a Born-Markov approximation, one recovers the standard equation 
\begin{equation}
	\partial_t a_r(t) = -\frac{\gamma_r(t)}{2}  a_r(t) - \sqrt{\gamma_r(t)}e^{\theta_r(t)} f_{\rm in, r}(t), \nonumber
\end{equation} \label{Eq:AppGamma}
with the phase
\begin{equation}
	\theta_\sigma(t) = \arg [g_{\rm sp}^{(\sigma)}(t)] + \phi_{m=0}.
\end{equation}
The effective decay rate of the TLS into the waveguide reads
\begin{align} \label{Eq:Geff}
	\gamma_\sigma(t) & = \frac{2\pi}{N_x} |g_{\rm sp}^{(\sigma)}(t)|^2\sum_{k_x}  |u_{s_\sigma}(k_x)|^2 \delta(\omega_E(k_x) - \omega_0), \nonumber \\
	 & = \frac{ 2|u_{s_\sigma}|^2}{v_g/a}|g_{\rm sp}^{(\sigma)}(t)|^2.
\end{align}
Here, $u_{s_\sigma} \equiv u_{s_\sigma}(k_x = k_0)$, $v_g \equiv  v_g(k_x = k_0)$ and $\phi_{m=0} \equiv \phi_{m=0}(k_x = k_0)$, where 
the momentum $k_0$ is defined as $\omega_E(k_0) = \omega_0$, i.e.~the momentum at which the frequency of the TLS crosses the dispersion relation of the edge modes.
Note that the factor $2$ in the second line of Eq.~\eqref{Eq:Geff} comes from the distance of $2a$ between two unit cells in the Kagome lattice. For example, in cases where only the atoms $B$ and $C$ are excited along the straight edges, i.e.~truly 1D limit [cf.~Fig.~\ref{Fig:EdgeState} (b)], $u_{s_\sigma} = 1/\sqrt{2}$ and $\gamma = a|g_{\rm sp}|^2/v_g$, as expected in a 1D unidirectional waveguide.
Finally, the incoming field reads
\begin{align}
	& f_{\rm in, r}(t) = f_{\rm out, e}(t-\tau_{er})e^{i\phi_{er}} \\
	& = [f_{\rm in, e}(t - \tau_{er}) + \sqrt{\gamma_e(t - \tau_{er})}e^{-i\theta_e(t-\tau_{er})}a_e(t-\tau_{er})]e^{i\phi_{er}}, \nonumber
\end{align}
with the propagation time and phase
\begin{equation}
	\tau_{er} = 2(j_r - j_e)a/v_g, 
	\qquad
	\phi_{er} = 2k_0 (j_r - j_e)a.
\end{equation}
This result is expected from the input-output formalism.

The role of the penetration depth $\xi \equiv \xi(k_x=k_0)$ is implicitly included in the coefficients $u_{s_\sigma}$ as the normalization constraint imposes 
\begin{equation}
	\sum_{s,m} |u_{s_\sigma}^2| e^{-\frac{2ma}{\xi}} \equiv 1.
\end{equation}
As expected, as the penetration depth increases, the strength at which the TLS couples to the edge state decreases.

\end{document}